\magnification \magstep 1
\def\today{\ifcase\month\or
  January\or February\or March\or April\or May\or June\or
  July\or August\or September\or October\or November\or December\fi

  \space\number\day, \number\year}
\hsize 6 true in
\vsize 8.5 true in
\input amssym.def
\input amssym.tex
\openup 1 \jot
\centerline { MASTER EQUATIONS AND MAJORANA SPINORS }
\vskip 1.5 cm
\centerline {G.W. GIBBONS}
\centerline {D.A.M.T.P.}
\centerline {University of Cambridge}
\centerline {Silver Street}
\centerline {Cambridge CB3 9EW}
\centerline {U.K.}
\vskip 1.5cm
\centerline {\rm \today}
\vskip 1.5cm
\centerline  {\bf ABSTRACT}                    
\tenrm
{\narrower \narrower. \smallskip
I describe how the  states of  a discrete automata
with $p$ sites, each of which may be off or on, 
can be 
represented as Majorana spinors associated to 
a spacetime with signature $(p,p)$. Some ideas about
the quantization of such systems are discussed and the relationship
to some  unconventional formulations
and generalizations  of quantum mechanics,
particularly  Jordan's spinorial quantum mechanics are pointed out.
A connection is made to  the problem of time and the complex numbers
in quantum gravity. }

\beginsection Introduction

Andre Trautman has had, throughout his career, the 
happy knack of realizing long ahead of others what will 
ultimately turn out to be just the right language and concepts
needed to most elegantly and economically encapsulate 
our emerging ideas about the physical world. Such languages must
be both precise and universal, easily understood and yet sufficiently
subtle as to suggest new avenues for exploration and exploitation. 
Often called geometrical, they may in fact incorporate
profound algebraic insights as well. 

I first became aware of Andre's talent for clear structured exposition
of a physical theory  on reading as a student his 1964 Brandeis Lectures, 
appropriately entitled " Foundations and Current Problems of General
Relativity". A later notable example of his prescience was his insight 
that fibre-bundle theory would eventually prove 
to be an everyday tool for physicists. 
More recently he has been exploring the world of spinors and 
geometrical algebra. In this article I wish,
among other things,
 to indicate how 
this might just turn out to be the appropriate 
language to describe such disparate subjects as the 
spread of forest fires, neural nets and the problem
of memory or the development of
avalanches, just as it has already shown its worth 
in statistical mechanics.

My own interest was first aroused by some ideas of 't Hooft's on cellular
automata and quantum mechanics [1,2]. During a stay in Utrecht I tried to understand what was 
the connection bwetween "spin models" and spinors which was  so 
effectively exploited by by Onsager 50 years ago in his solution of the 
Ising model. Since then I have been struck by how often the same ideas
seem to crop up in many other areas of physics of which 
I have only a superficial knowledge. Strange analogues of 
quantum mechanics, quantum field theory and the Dirac equation 
seem  to arise all over the place. Clearly there  is
something universal going on here. 
What follows is my attempt to understand what that might be.
I do not claim that I am alone in realizing this connection
(Dubois-Violette, for example, has described  similar things [3])
but my emphasis on  the essential  reality (in the mathematical
sense) of the construction and my attempts (in the final section) to relate
the discussion to some ideas in quantum gravity is perhaps new. 
In any event  I dedicate it to Andre hoping that he will find
my exposition of interest
and that it may perhaps provoke him to look further 
if he hasn't already done so.

\beginsection 1. Master Equations and Partition Functions

The common theme of many of the applications I alluded to is a 
Markov process in which the time development of a "probablity vector"
${\bf p}= (p_1,p_2, \dots ,p_n)$ evolves according to a "Master Equation"
of the form
$$
{d p_i \over dt} = \sum _{j\ne i}  \Bigl ( -w_{ij} p_i + w_{ij} p_j \Bigr )
\eqno(1.1)
$$
$$
= \sum _j w_{ij} p_j,
\eqno(1.2)
$$
where the "transition rates" $w_{ij}$ are positive and constrained
by the requirement that
$$
\sum _i p_i=1, \eqno(1.3)
$$
so that for example  
$$w_{11}= -w_{12}-w_{13} \dots - w_{1n}. \eqno(1.4)
$$
Geometrically speaking (1.3) tells us that the probability vector 
lies on the $(n-1)$-simplex  
$\Sigma_{n-1}$ in the positive orthant ${\Bbb R}^n _+$ of ${\Bbb R}^n$ \footnote {*}{ By ${\Bbb R}^n$, I mean the standard $n$-dimensional vector space
equipped with its positive definite inner product. 
I shall not, therefore, distinguish between vectors and covectors.
All indices will therefore be lowered.}  
satisfying 
$$
{\bf i} . {\bf p}=1
\eqno(1.5)
$$
where ${\bf i}=(1,1,\dots 1)$ is the un-normalized 
completely ignorant probability vector obtained by taking all 
{\it a priori}
probablities to be equal. The set of probability vectors, 
also called states,
is convex. The extreme points of the state space $\Sigma _{n-1}$,
may be are called "pure states", and correspond to the vertices of 
the simplex,
that is to the unit basis vectors ${\bf e}_i$ in ${\Bbb R}^n$
and the remaining, non-extreme states are called "mixed".   
Note that the language of pure and mixed states is as applicable to 
the classical systems being considered here as it is to 
quantum mechanical systems. Indeed according to one view point 
what makes the present set up classical is precisely the fact that the 
space of mixed states $\cal M$ {\sl is} a simplex [4,5,6]
and the space of pure states $\cal P$ are its vertices.  
  
The solution of (1.1) may be written as
as 
$$
p_i(t) = \exp (t w_{ij})p _j(0)
\eqno(1.6)
$$
$$
= \$_{ij} p_j(0),
\eqno(1.7)$$
where  $\$ = \exp {t w}$ is a stochastic matrix, i.e a square  matrix
all of whose rows sum to unity.

The stationary, i.e time-independent, states, or 
equilibria have probabilities $p_i^{\rm equ}$. In thermodynamic 
applications each pure state ${\bf e}_i$ is assigned an energy $E_i$
one has at temperature $T= \beta ^{-1}$,
$$
p_i^{\rm equ}= \exp (-\beta E_i) /Z(\beta)
\eqno(1.8)$$
where:
$$
Z(\beta) = \sum _i \exp (-\beta E_i).
\eqno(1.9)$$
is the partition function.  Given the energies $E_i$
the challenge is  to calculate the partition function $Z(\beta)$.
This was achieved by Onsager [7] using 
Pauli matrices including the {\sl pure iamaginary} matrix $\sigma_2$ for the two-dimnsional lattice Ising model. 
I shall argue shortly that, despite appearances and 
as befits the
purely  classical  nature of the problem, these spinorial techniques
actually involve the {\sl real}, i.e. Majorana spinors of a Kleinian spacetime
with as many time coordinates as it has space coordinates. 

More generally given an observable ${\bf O}$ which takes values
$ O_n$ , its expectation value $\langle {\bf O }\rangle$ at time $t$
is given by
$$
\langle {\bf O} \rangle = \sum _i p_i O_i.
\eqno(1.10)$$

One may think of the classical (commuting)
observables as matrices acting on ${\Bbb R}^n$
which happen to be diagonal in the basis provided by the pure states. 
One may then re-write (1.10) as
$$
\langle {\bf O} \rangle= {\bf i} . {\bf O  p}.
\eqno(1.11)$$

\beginsection 2 Quantizing Discrete Automata

I have already commented upon the superfical resemblance
of this classical  formalism to  standard quantum mechanics.
If we really were dealing with quantum mechanics,
then the corresponding configuration space $\cal Q$ would be just the 
the discrete set of $n$ points ${\cal Q}_n= \{{\bf e}_i\}$ 
and the Hilbert space $\cal H$ would be $L^2 ( {\cal Q}_n, {\Bbb C} ) $
which may think of as the standard $n$-dimensional
Hermitean vector space ${\Bbb C}^n= {\Bbb R} ^n \otimes {\Bbb C} $
with orthonomal basis given by  $\{{\bf e}_i\}$. 

Thus even though  the ({\sl linear}) equation (1.2) is like the ({\sl linear})
Schr\"odinger equation we face the difficulty that 

\medbreak \item {(i)} The probablity vector ${\bf p}$ has 
components which are real and positive. 

\medbreak \item {(ii)} 
The probablity vector ${\bf p}$ is normalized using the 
$L^1$ norm (1.4) rather than the usual $L^2$ norm,

\noindent One might consider, as does Wheeler in considering thesis topics 
[9],
 passing to real quantum mechanics [8] by taking a square root,
i.e. introducing the vector
$$
{\bf \psi} = \bigl ( (p_1)^{ 1 \over 2}, (p_2)^{ 1 \over 2},\dots (p_n)^{ 1 \over 2} \bigr ).
\eqno(2.1)
$$
The normalization  condition (1.4) now tells us that $\psi$ lies on the unit
$n$-sphere $S^{n-1} \subset {\Bbb R}^n$. However now the equation
of motion (1.2) becomes {\sl non-linear}. Moreover there is some ambiguity
in taking the square roots in (2.1). If we take them all to be positive
we map the simplex $\Sigma _{n-1}$ onto the positive orthant of the $(n-1)$-sphere or projectively speaking, the interior of 
an $(n-1)$-simplex sitting inside real projective space ${\Bbb R} {\Bbb P}^{n-1}$. In real quantum mechanics one usually thinks of the pure states
as all real vectors $\psi$ except that one identifies 
those which differ by a non-zero real multiple. Thus the  the pure
states
roam all over ${\Bbb R} {\Bbb P}^{n-1}$ . To restrict them to a part
of ${\Bbb R} {\Bbb P}^{n-1}$ would in effect put a restriction on the 
Superposition Principle : only postive real combinations would be allowed.    

The idea pursued by 't Hooft is different. He considers 

\medbreak \item {(i)} The  evolution to be reversible
and time to be discrete

\medbreak \item {(ii)} Each site to be occupied with certainty.  

\noindent Thus we have a classical reversible discrete automaton whose time evolution is given by iterating a permutaion matrix $U : {\cal Q}_n \rightarrow {\cal Q}_n$ which permutes the basis elements $\{ {\bf e}_i \}$ of ${\Bbb R}^n$.
After $N$ steps we obtain $U^N$. Clearly $U^N$ is a special case
of an orthogonal matrix
$$
U^N \subset SO(n; {\Bbb R})
\eqno (2.2)
$$
or  indeed a unitary matrix
$$
U^N \subset U(n; {\Bbb C}).
\eqno (2.3)
$$
't Hooft's idea is to find a Hermitean matrix $H$ such that
$$
U^N= \exp ( -iN H).
\eqno(2.4)
$$
The Hermitean matrix $H$ is to be thought of as the Hamiltonian
of the quantum mechanical
system associated to the original classical automaton. 

Quite aside from the ambiguity in taking the logarithm
in order to obtain $H$,
there are two rather puzzling things about this.

\medbreak \item {(i)}
Usually we only consider classical first order equations
governing the motion in classical state space or phase space  $\cal P$
which is even dimensional.
The motion in configuration space $\cal Q$ is usual governed by 
second order equations. Thus we might have tried to identify
$\cal P$ with the discrete set of states $\{ {\bf e}_i \}$.

\medbreak \item {(ii)} Usually the quantum mechanical Hilbert space
$\cal H$ is not given by $L^2 ({\cal P})$, rather one seeks a 
prescription (e.g a "polarization") for cutting $L^2 ({\cal P})$ down to 
a  Hilbert space of functions depending upon half as many variables.
The standard continuous case is of course when the phase
space is the tangent bundle of the configuration space
${\cal P} = T^* ( {\cal Q})$.

This  suggests to me that one should restrict attention 
to the case when the number of discrete states $n$ is even. One could
then seek to embed the time evolution operator 
in $U({ n \over 2}; {\Bbb C})$, 
$$
U^N \subset U({ n \over 2}; {\Bbb C}).
\eqno(2.5)
$$
 In other words one should  
try to endow the original real vector space ${\Bbb R}^n$ 
with a complex structure $J$ thus allowing us to regard it as
${\Bbb C}^{ n \over 2}$. Acting on the states one would have:
$$
{\bf e}_i \rightarrow J({\bf e}_i)
\eqno (2.6)
$$
such that $J^2= -1$ and moreover one would require
that this action commutes with the time evolution:
$$
UJ-JU=0.
\eqno (2.7)
$$  
The complex structure $J$ would then play  the analogous
role to that of 
a "polarization" in the geometrical quantiztion of a symplectic manifold.  

An interesting and different example where one passes from a 
classical to a quantum 
desription of a discrete system arises in the the theory of 
quantum computation [11]. Classically a single bit of information is carried
by two-state system. Quantum mechanically a qbit is carried by a two-state
quantum sytem. However in that case it is physically more
realistic to think of
the classical two state system arising from the quantum-mechanical
two-state system in the limit that all quantum coherence is lost
rather than thinking of the quantum system as arising from
the "quantization of the classical system " . Thus
classically one has a $2\times2$ diagonal density matrix 
whose entries are $ p_\downarrow $ and $+ p_\uparrow $ say.
Thus the remarks of the last 
paragraph do not apply.

\beginsection 3 Site Models

So far I have said nothing about what the probabilities $p_i$
are supposed to be for. Moreover the systems we have considered have no 
"spatial" structure and so notions like locality are not defined.
In many applications one considers a certain number, $p$
 of sites which may be occupied or not. Typical examples
are "classical spins" which may be "up" or "down" in the Ising model
or trees in a forest which may be either "green" or "burnt out". The total number
of pure states $n$ is therfore $2^p$. This is exactly the dimension
of the Grassmann algebra $\Lambda ^* ({\Bbb R} ^p)$ over $ {\Bbb R} ^p$
and the connection arises as follows. We have
$$
\Lambda ^* ({\Bbb R} ^p)= \sum _{q=0} ^ { q=p} 
\oplus \Lambda ^p ({\Bbb R} ^p).
\eqno(3.1)
$$
where $\Lambda ^p ({\Bbb R} ^p)$ are the $q$-forms on $ {\Bbb R} ^p$.
The space of $q$-forms is spanned by the simple $q$-forms. These correspond to the $q$-planes through the origin of ${\Bbb R} ^p$ which contain 
$q$ distinct basis vectors. There are ${p! \over q! (p-q)!}$ of these
and they correspond to the ways of choosing $q$ occupied sites. Thus
for exmple $\Lambda ^0 ({\Bbb R} ^p) \equiv {\Bbb R}$ correponds to no sites being occupied and $\Lambda ^p ({\Bbb R} ^p) \equiv {\Bbb R}$ 
correponds to all sites being occupied. Hodge duality acting on $\Lambda ^* ({\Bbb R} ^p)$
reverses the state of occupation of the sites.

Thus if 
we think of a single site ${\bf e}_1$ and $p_\downarrow $ is the probability of 
it  being empty and $p_\uparrow$ is the probability of 
it  being full we have the probablity vector
$$ 
{\bf p} = p_\downarrow + p_\uparrow {\bf e}_1.
\eqno( 3.2)
$$
If two sites are involved we have 
$$
{\bf p} = p_{\downarrow \downarrow} + p_{\uparrow \downarrow }{\bf e}_1
+  p_{\downarrow \uparrow }{\bf e}_2
+ p_{\uparrow \uparrow } {\bf e}_1 \wedge {\bf e}_2.
\eqno(3.3)
$$
As an example one could take Empedocles's theory of the four elements,
fire, earth air and water based on the two pairs of contrary attributes
hot/cold and dry/wet. Ramon Lull's {\it Ars Magna} 
[10] is essentially the generalization to more than 
two sites.

These last two examples look, on the face of it,
 neither quantum mechanical nor spinorial
but let's  define real creation and anhillation operators
$a_a$ and $a^\dagger _a$ where the index $a$ taken from the beginning
of the latin alphabet run from $1 $ to $p$ by their action on a general
probability vector ${\bf p} \in \Lambda ^* ({\Bbb R} ^p)$ by
$$
a_a^\dagger {\bf p} = {\bf e}_a \wedge {\bf p}
\eqno( 3.4)
$$
and
$$
a_a {\bf p} = i_{{\bf e}_a} {\bf p},
\eqno(3.5)
$$
where $\wedge$ is the exterior product and $i$ denotes the interior product.
It follows that
$$
a_a^\dagger a_b + a_b a_a^\dagger = \delta _{ab}
\eqno(3.6)
$$

$$
a_a^\dagger a^\dagger _b + a^\dagger _b a_a^\dagger = 0
\eqno(3.7)
$$ 
and
$$
a_a a_b + a_b a_a = 0.
\eqno(3.8)
$$
In fact it is straightforward to check that, with respect to the euclidean inner product induced on $\Lambda ^* ({\Bbb R} ^p)$ form the euclidean
inner product on ${\Bbb R} ^p$, that $a_a$ and $a_a ^\dagger$ are adjoints
or in the natural basis transposes:
$$
a^\dagger _a = (a_a)^t.
\eqno(3.9)
$$

If one changes  basis by introducing   
$$
\gamma _{\pm a}= a_a \pm a_a ^ \dagger
\eqno(3.10)
$$
one has
$$
\gamma _{\pm a} \gamma _{\pm b} + \gamma _{\pm b} \gamma _{\pm a}
= \pm 2 \delta _{ab}
\eqno(3.11)
$$
and 
$$
\gamma _{+ a} \gamma _{- b} + \gamma _{- b} \gamma _{+ a}
= 0
\eqno(3.12)
$$
Thus one sees that the annihillation and creation
operators $\{ a_a, a_a^\dagger \}$ generate the real Clifford algebra
$${\rm Cliff} ({\Bbb R} ^{p,p}) \equiv {\Bbb R} (2^p)$$
where $ {\Bbb R} (2^p)$ is the algebra of real $2^p \times S^p$ matrices. 
In other words we may think of the probability vectors $\bf p$ as
being Majorana spinors for $SO(p,p; {\Bbb R})$, hence my title.
The space of mixed states $\cal M$ is thus
the set of Majorana spinors  whose components are 
non-negative and which sum to unity. The set of pure states 
$\cal P$ are those spinors with one non-vanishing component
equal to unity. Unfortunately they do not in general correpsond to 
what are called
"pure spinors". 

  Note that the Clifford group itself is the general linear group
$GL(2^p; {\Bbb R})$. The annihillation and creation
operators $\{ a_a, a_a^\dagger \}$ themselves are associated with 
lightlike directions. The volume element $\eta \in \Lambda ^p ({\Bbb R})$
is given by
$$
\eta = \prod _{a=1}^ {a= p} \bigl ( 1- 2 a^\dagger _a a _a \bigr ).
\eqno(3.13)$$
Thus $\eta = 1$ for states with an even number of occupied sites and $\eta=-1$ for states with an odd number of occupied sites.
It follows that the direct sum decomposition:
$$
\Lambda ^ * ( {\Bbb R} ^p)= \Lambda ^ {\rm even} ( {\Bbb R} ^p) \oplus \Lambda ^ {\rm odd} ( {\Bbb R} ^p)
\eqno(3.15)
$$
corresponds to the direct sum decomposition of the space of Majorana spinors
into Weyl and anti-Weyl spinors. Weyl spinors have an even number of occupied sites and anti-Weyl spinors an odd number of occupied
sites.

\beginsection Spinors, the Light Cone and Jordan Quantum Mechanics

In view of the connection emphasised by Penrose between
2-component Weyl spinors of $SO(3,1)$, the lightcone of
four-dimensional Minkowski spacetime and the quantum mechanics of spin 
It seems appropriate, while considering quantum mechanics and spinors,
to recall a generalization of the usual quantum mechanical formalism due
to Jordan which dispenses with wave functions
and takes density matrices as being fundamental.
As remarked by Townsend, [12] this is especially 
intriguing because of Hawking's well known 
suggestion 
that quantum coherence may be lost in quantum gravity
due to black hole effects.    
Jordan  requires of his "observables" $ \rho $
 they satisfy the axioms of a real commutative but non-associative
algebra ${\cal A} = \{\rho, \bullet \}$ now called a Jordan Algebra. 
The irreducible finite-dimensional 
algebras were classified by Jordan, von Neuman and  Wigner [13]. 
There are four series and one exceptional case:

\medskip \item {(i)} ${\rm J }( {\Bbb R} ^n)$
\medskip \item {(ii)} $H^{\Bbb R}_n$
\medskip \item {(iii)} $H^ {\Bbb C}_n$
\medskip \item {(iv)} $H ^{\Bbb H}_n$
\medskip \item {(v)} $H^{\Bbb O}_3$

Of these $H^{\Bbb R}_n,H^ {\Bbb C}_n,H ^{\Bbb H}_n$ consist of $n \times n$
real symmetric, Hermitean and quaternionic
Hermitean matrices respectively , corresponding to quantum mechanics over
the fields ${\Bbb R}, {\Bbb C} , {\Bbb H}$ for which the product $\bullet$
is given by
$$
\rho_1 \bullet \rho_2 = { 1\over 2} \Bigl ( \rho _1 \rho _2 + \rho _2 \rho _1 \Bigr ).
\eqno (4.1)
$$
The exceptional case $H^{\Bbb O}_3$ is 
related to $3\times3$ the octonionic matrices.  The $(n+1)$-dimensional  sequence ${\rm J }( {\Bbb R} ^n)$
is based on 
Clifford multiplication associated to  ${\Bbb R}^n$.
If $\{ 1, \gamma _i \}$ is an orthormal basis for the Clifford algebra
${\rm Cliff}( {\Bbb R}^n)$ a general mixed state may be written as
$$
\rho =  { 1\over 2}\Bigl ( x_0 + x_i \gamma _i \Bigr )
\eqno(4.2)
$$
and the Jordan product satisfies
$$
\gamma _i \bullet \gamma_j= \delta _{ij}.
\eqno(4.3)
$$
 Note that the non-associative algebra ${\rm J }( {\Bbb R} ^n)$ is not a subalgebra of 
 associative algebra ${\rm Cliff}( {\Bbb R}^n)$ because they use a different multiplication law.
However given a matrix representation  of the Clifford algebra 
one obtains a matrix represention of the Jordan algebra using (4.1).

Note that as algebra's Jordan algebras are necessarily real
and indeed one may regard this
as a  part of the motivation: to deal only with real observables
quantities. However it may happen of course that the  spinorial
Jordan  algebras
${\rm J} ({\Bbb R}^n)$ based on the real Cliford algebra ${\rm Cliff} ({\Bbb R}) ^n )$ are isomorphic to algebras of complex valued matrices.
Whether or not this is true depends upon the dimension $n$.
This happens  for the case $n=3$, which may be said in some way
to account for the sucess of complex methods in 
four-dimensional general relativity.

Let us  return to the interpretation of spinorial Jordan algebras. 
If $n_i= { x_i \over (x_k x_k ) ^ {1 \over 2} }$
then provided one sets $x_0= 1$
every non-trivial mixed state $\rho$ may be expressed uniquely in terms
 of two pure states which are associated to two
primitive idempotents or projection operators $E_\pm = { 1 \over 2 } \bigl( 1 \pm n_i \gamma_i \bigr )$
$$
\rho = p E_+ + (1-p) E_-. \eqno(4.4)
$$
The idempotents $E^\pm$ satisfy
$$
E_\pm ^2 = E_\pm . \eqno(4.5)
$$
and
$$
E_\pm \bullet E_ \mp =0 . \eqno(4.6) 
$$
   
Since $|x|= |p- { 1\over 2}|$, and one wishes to interpret $p$ as a 
probability to be in the pure state associated to $E_+$,
observables are confined to lie within
or on the surface of a ball 
of unit {\sl diameter}  in ${\Bbb R}^n$. The mixed states $\cal M$
lie in the interior and the pure states $\cal P$, which correspond to idempotents
of the algebra, lie on the unit $(n-1)$ sphere. 
$$
x_i x_i =1.
\eqno (4.7)
$$
Geometrically speaking, a  general mixed state lies on a unique  
diameter of the ball.
The ends of which are the pure states. The probabilities
to be in the two pure states are given by the distances to the 
endpoints of the diameter.

Clearly we may think of the set of pure states 
of the Jordan algebra ${\rm J} ({\Bbb R}^n)$ as the light cone
of $(n+1)$-dimensional Minkowski spacetime ${\Bbb R}^{n,1}$.
There are four  special cases:
\medskip \item {(i)} ${\rm J} ({\Bbb R}) \equiv {\rm Cliff} ( {\Bbb R}) \equiv {\Bbb E} \equiv {\Bbb R} \oplus {\Bbb R}$,
\medskip \item {(ii)} ${\rm J} ({\Bbb R}^2) \equiv H_2 ^ {\Bbb R}$,
\medskip \item {(iii)}${\rm J} ({\Bbb R}^3) \equiv H_2 ^ {\Bbb C}$,
\medskip \item {(iv)}${\rm J} ({\Bbb R}^5) \equiv H_2 ^ {\Bbb C}$.

 Case (i) is that of a single  a classical spin. 
The symbol $\Bbb E$ denotes the double or hypercomplex numbers.
As an algebra
it is reducible. The space of pure states is $S^0 \equiv {\Bbb Z_2}$.

Case (iii)  coincides with the smallest possible
 non-trivial standard quantum mechanical
system  having  just two states with Hilbert space ${\Bbb C}^2$. The 
space of pure states is $S^2 \equiv {\Bbb C} {\Bbb P} ^1$.

Case (ii) is the real quantum mechanical 
version of this  with Hilbert space 
${\Bbb R}^2$. 
The space of pure states is $S^ 1 \equiv {\Bbb R} {\Bbb P} ^1$.

Case (iv) corresponds to quarternionic quantum mechanics 
with Hilbert space ${\Bbb H}^2$. 
The space of pure states is $S^ 4 \equiv {\Bbb H} {\Bbb P} ^1$.

The other cases  differ radically from standard quantum,
or non-standard
mechanics. In general  there is no simple correspondence beween
light rays or pure states and spinors and  the spinors
are not in general Majorana. However, as we shall see shortly,
one may also discuss master equations in the context of 
spinorial Jordan mechanics.

It is convenient to consider only observables which have no
component along the direction of the unit matrix. 
Then, up to scale, they may be identified with points on 
the sphere with unit diameter.

If we have a matrix representation  we may express the expectation value of 
an observable $a$ in a state $\rho$ as 
$$
\langle a \rangle ={2 \over \nu } {\rm Tr} \rho a = 2a_i x_i,
\eqno (4.7.) $$
where $\nu$ is the dimension of the representation of the Clifford algebra, ii.e. $2^{[{n \over 2}]}$. We may write the expectation value as
$$
\langle a \rangle= p(a_i n_i) + (1-p) (-a_i n_i), \eqno(4.8)
$$
where $p$ and $1-p$ are the probabilities of being
in the two pure states at the two ends $\pm n_i$ 
of the diameter whose direction is given by the unit vector
$n_i$. 

We can associate an observable $a$ with the 
real function $a(y)=2a_i y_i$ on the $(n-1)$ sphere 
of points  $\{y_i| y_k y_k = {1 \over 4} \}$. 
Similarly the mixed state $\rho$ gives rise
to the non-negative function $P(y)=  1  +4 x_iy_i$
with mean value $1$. 
One may then interpret $P(y)$ as a sort of classical probability 
distribution
and we have
$$
\langle a \rangle= { n \over 2}  { \int _{S^{n-1}} \mu  \rho (y) a(y) }  /  \int _{S^{n-1}}\mu,
\eqno (4.9.)
$$
where $\mu$ is the volume element on the $(n-1)$-sphere of 
radius $1 \over 2$. If $n=2$ the probability distribution
interpretation is consistent. It is a special case
of the general quantum mechanical discussion on ${\Bbb C} {\Bbb P} ^n$
[14].

The difficulty with spinorial Jordan  mechanics 
emphassised by Townsend is the introduction of 
the analogue of a Shr\"odinger  equation.
In standard quantum mechanics, just as in classical  mechanics,
every observable generates a flow on the space $\cal P$ of pure states
which extends to a flow on the space of mixed states $\cal M$. 
One may alternatively postulate,
particularly in the case of open systems, 
that the fundamental equation of motion is
a generalized master equation or flow on the space of mixed states
$\cal M$ which need not necessarily be induced from a 
flow on the pure states. 
This is essentally Hawking's viewpoint: he takes
${\cal M} = H_n^ {\Bbb C} $, the space of density matrices in standard
quantum mechanics. The finite (or infinite ) 
time transformations correspond to his linear $\$$ map
called the super-scattering matrix:
$$
\rho \rightarrow \$\rho. \eqno(4.10)
$$
One says that the $\$$ matrix factorizes if this map may be
written as  
$$
\rho \rightarrow S \rho S^\dagger \eqno(4.11) 
$$
for a unitary  $S$-matrix $S \subset U(n;{\Bbb C})$.    
Note that that if one wishes to
retain the usual interpretation of probabilities
relating them to the proportions of the outcomes
of independent experiments one must have a {\sl linear} 
master equation.  

In the case of the  standard time evolution  
for finite dimensional quantum mechanical systems with $n$ states 
one uses the fact that ${\cal P} \equiv {\Bbb C} {\Bbb P}^{n-1}$
is a symplectic manifold
and the flow is  Hamiltonian. The set of observables 
generate the group $U(n;{\Bbb C})$ which acts transitively on ${\Bbb C} {\Bbb P}^{n-1}$. In the case of  spinorial Jordan mechanics an observable $a $
certainly gives rise to  real valued function (i.e. $a_ix_i$)
but  
 ${\cal P} \equiv S^n$ is only symplectic for for $n=2$.
One way to get a flow would be to take the gradient flow 
with respect to the  round metric on  $S^{n-1}$ . 
Each such function is a conformal Killing potential
and so 
would generate a one parameter subgroup of the 
conformal isometries of $S^{n-1}$. However these do not act linearly
 on the space of mixed states ${\cal M} \equiv B^n$.
It seems therefore that in spinorial Jordan mechanics one must
give up the idea that a single observables can generate the dynamics.

An alternative approach to finding an equation of motion is to use {\sl two} obervables $a$ and $b$ [12].
The analogue of the Heisenberg equations of motion for 
a density matrix $\rho$ 
is then taken to be
$$
{\dot \rho} + \{a, \rho, b \}=0, \eqno(4.12)
$$
where the {\it associator} of the three elements $\{a,\rho, b\}$
is defined by
$$
\{a,\rho, b\}= (a \bullet \rho)\bullet b -a \bullet (\rho \bullet b).
\eqno (4.13)
$$
If $a,\rho,b$ are matrices one has
$$
\{a,\rho, b\}= { 1 \over 4} \bigl [ \bigl [ b,a \bigr ], \rho  \bigr ],
\eqno (4.14)
$$
where $\bigl [  b, a \bigr ]$ is the ordinary matrix commutator.
In ordinary quantum mechanics we can always find  
$a$ and $b$ such that the Hamiltonian $H$ satisfies
$$
H= { i \over 4} \bigl [a,b \bigr ]
\eqno (4.15)
$$
and so (4.16) reduces to Heisenberg's equation:
$$
{\dot \rho} + i \bigl [H,\rho \bigr ]=0.
\eqno(.10)
$$

A simple calculation shows that (4.12) becomes in components
$$
{\dot x}_i = ( b_i a_j-a_i b_j)x_j
\eqno(4.17)
$$
This is just a rotation in the $2$-plane spanned by the $n$-vectors
whose components are $a_i$ and $b_i$. It certainly does not 
correspond to a gradient flow. However if (4.12) is
generalized to include a sum of associators of pairs
of elements one would obtain a general element of $so(n;{\Bbb R})$.
Thus in spinorial Jordan mechanics the natural time evolution law 
seems corresponds to rigid
rotations of the space ${\cal P} \equiv S^{n-1}$ of pure states.  

Following Hawking one might introduce a linear map 
$\$ : B^ n \rightarrow B^n$ which fixes the totally ignorant state 
$ \rho = { 1 \over 2} $,
i.e which fixes the origin. Such maps would be the product of
a purity preserving rotation combined with a purity
reducing contraction. The invertible $\$$ maps then
would be the rotations generated as above.

The conclusion seems to be that in general Jordan's algebraic
formulation of Quantum Mechanics, in particular
the spinorial version,  fails the test of 
providing a satisfactory time evolution law unless
it coincides with the conventional case.

\beginsection Conclusion

What I have attempted to do in this article is to show that 
spinorial techniques using  or Majorana spinors provide
 a convenient language for describing discrete classical systems 
such as disrete or cellular automata. The observation 
that spinor techniques may be useful is itself is not
new. However what is often not greatly stressed is that one only needs
{\sl real} numbers. 
Complex numbers only come in when one
considers quantum systems and their introduction is not
as completely trivial as is sometimes supposed. In conventional
quantum mechanics the complex numbers are needed when considering
 time evolution. This problem  of introducung  a time function
and a complex structure into quantum
mechanics  becomes especially acute in 
quantum cosmology [15,16] where the consensus appears to be moving
towards the idea that {\sl both} are approximate concepts. 

It is therefore of  interest to see 
what form this problem takes in generalizations
of conventional quantum mechanics.
An illuminating example is provided by  Jordan's  spinorial 
algebraic version of  quantum mechanics. We found that 
although there is a striking connection between
the light cone of $n+1$-dimensional Minkowski spacetime
and spinors. the enterprize founders on the problem of time evolution
unless it coincides with the orthodox  spin $1 \over2 $
quantum case. 
However, as I have argued elsewhere [15,16], this does not necessarily mean
 that the complex numbers
of the spin group $Spin(3,1) \cong SL(2, {\Bbb C})$ are to be identified
with the complex numbers of quantum mechanics. Indeed one could formulate
all of classical  physics in terms of Majorana spinors and stll have to introduce
complex numbers to pass to the quantum theory [15,17]. 
Whilst thay are undoubtedly useful,
Weyl spinors are in no way obligatory.

\beginsection References

\medskip \item {[1]} G 't Hooft { \sl J Stat Phys} {\bf 53} (1988) 323-344

\medskip \item {[2]} G 't Hooft {\sl Nucl Phys } {\bf B 342} (1990) 471-485 

\medbreak \item {[3]} M Dubois-Violette : Complex structures and the Elie
Cartan approach to the theory of Spinors  in {\sl Spinors, Twistors, Clifford Algebras and Quantum Deformations} 
eds. Z. Oziewicz et al., Kluwer, Amsterdam (1993)

\medskip \item {[4]} B Mielnik {\sl Commun Math Phys} {\bf 9} (1968) 55-80

\medskip \item {[5]} B Mielnik {\sl Commun Math Phys} {\bf 15} (1969) 1-46

\medskip \item {[6]} B Mielnik {\sl Commun Math Phys} {\bf 37} (1974) 221-256

\medskip \item {[7]} L Onsager {\sl Phys Rev} {\bf 65} (1944) 117-147

\medskip \item {[8]} E C G Stueckelberg {\sl Helv Phys Acta} {\bf 33} (1960) 
727-752

\medskip \item {[9]} J A Wheeler in {\sl Directions in General Relativity: vol I} edited by  B L Hu, M P Ryan and C V Visveshwara, Cambridge University Press (1993) 

\medskip \item {[10]} M Gardner {\sl Logic Machine, Diagrams and Boolean Algebra} Dover Publications, New York (1968)

\medskip \item {[11]} D Deutsch {\sl Proc R Soc Lond} {\bf A 400} (1985) 97-117 

\medskip \item {[12]} P K Townsend, The Jordan Formulation of Quantum Mechaincs in {\sl Supersymmetry, Supergravity and Related Topics} edied by F del
Aguila, J A Azc\'arraga and L E Ib\'a\~nez, World Scientific (1985)

\medskip \item {[13]} P Jordan, J v Neuman and E Wigner {\sl Annals of Mathematics} {\bf 35} (1934) 29-64

\medbreak \item {[14]} G W Gibbons {\sl  J Geom and Phys} {\bf
8} 147--162 (1992)

\medbreak \item {[15]} G W Gibbons and H-J Pohle {\sl Nucl. Phys.} {\bf B 410} 117-142 (1993)

\medbreak \item {[16]} G W Gibbons {\sl  
Int. J. Mod. Phys. } {\bf D 3} 61-70 (1994)

\medbreak \item {[17]} G W Gibbons 
The Kummer Configuration and the Geometry of Majorana Spinors 
in {\sl Spinors, Twistors, Clifford Algebras and Quantum Deformations} 
eds. Z. Oziewicz et al., Kluwer, Amsterdam (1993)

\bye